\documentclass[twocolumn]{jpsj3}
\usepackage{ulem}
\usepackage{color}
\usepackage{braket}
\usepackage{cite}

\definecolor{mygreen}{rgb}{0.0,0.7,0.0}

\title{Finite-Temperature Variational Monte Carlo Method for Strongly Correlated Electron Systems}

\author{Kensaku Takai, Kota Ido, Takahiro Misawa, Youhei Yamaji, and Masatoshi Imada}
\inst{%
 Department of Applied Physics, University of Tokyo, 7-3-1 Hongo, Bunkyo, Tokyo 113-8656, Japan
}%

\abst{
A new computational method for  finite-temperature properties of strongly correlated electrons is proposed  by extending the variational Monte Carlo method originally developed for the ground state.
The method is based on the path integral in the imaginary-time formulation, starting from the infinite-temperature state that is well approximated by a small number of certain random initial states.
Lower temperatures are progressively reached by the imaginary-time evolution.
The algorithm follows the framework of the quantum transfer matrix and finite-temperature Lanczos methods, 
but we extends them to treat much larger system sizes without the negative sign problem
by optimizing the truncated Hilbert space
on the basis of the time-dependent variational principle (TDVP). 
This optimization algorithm is equivalent to the  stochastic reconfiguration (SR) method that has been frequently used for the ground state to optimally truncate the Hilbert space.
The obtained finite-temperature states allow an interpretation based on the thermal pure quantum (TPQ) state instead of the conventional canonical-ensemble average. 
Our method is tested for the one- and two-dimensional Hubbard models and its accuracy and efficiency are demonstrated. 
}

\begin{document}
\maketitle

%
\section{Introduction}
%
 Physics of strong electron correlations has been a central issue of condensed matter physics 
for decades. Many
open problems such as mechanisms of high-temperature superconductivity in copper oxides are still left as major challenges~\cite{RevModPhys.70.1039}.

To take steps forward, 
theoretical methods that are able to accurately describe the interplay between the kinetic motion of electrons and mutual Coulomb repulsions are required
beyond perturbation theory and mean-field descriptions.
Numerical algorithms developed for decades offer
reliable approaches for the purpose of treating this interplay.
Powerful numerical algorithms exist, such as the exact diagonalization (ED)~\cite{RevModPhys.66.763},
the quantum Monte Carlo (QMC) methods
\cite{PhysRevD.24.2278,0295-5075-8-7-014, JPSJ.58.3752, JPSJ.61.3331},
and the density matrix renormalization group (DMRG) method\cite{PhysRevB.48.10345} including the tensor network algorithm\cite{Orus2014117},
which have provided essentially exact solutions for effective Hamiltonians 
such as Hubbard-type model Hamiltonians.  
The cluster extension of dynamical mean-field theory (DMFT)~\cite{RevModPhys.77.1027,RevModPhys.78.865} also offers accurate numerical solutions if the large cluster size limit could be taken~\cite{PhysRevLett.106.030401}.

Despite their accuracy, existing numerical methods suffer from severe limitations in their applicability.
While the ED gives us, of course, exact results, it is only applicable to small finite-size clusters. 
Several QMC methods with high accuracy are applicable to larger systems.
However, the Hamiltonians that can be solved accurately with the QMC methods are limited owing to the so-called negative sign problem\cite{JPSJ.60.810}. 
The DMRG method is practically an exact method in one-dimensional configurations without any negative sign problem.
However, the accuracy of the DMRG method depends on the entanglement properties of the system:
The application of the DMRG method to two- and three-dimensional systems still remains challenging. 
The cluster extension of the DMFT method requires a large  computational cost that rapidly grows with the cluster size,
which hampers its application to the large cluster limit.

Continuous efforts have been made to overcome these limitations of the existing numerical algorithms.
When we focus on the ground-state properties of  strongly correlated electrons, several accurate variational wave function approaches have been developed.
An example is the tensor network approach\cite{Orus2014117}.
The tensor networks are a generalization of the matrix product wave functions employed in the DMRG algorithms,
which partially resolve the disadvantage of the DMRG method in its application to higher dimensional systems because larger entanglement can be handled. 
 Another example is the many-variable variational Monte Carlo (mVMC) method
\cite{JPSJ.77.114701}, which is an extension of traditional variational approaches
\cite{JPSJ.56.1490,Gros198953} by including the optimization of a systematically  large number of variational parameters, enabled by the development of an efficient optimization algorithm, such as the 
stochastic reconfiguration (SR) method~\cite{PhysRevB.64.024512}.
The mVMC method indeed allows the introduction of more than thousands of  variational parameters to 
reduce biases inherent in variational wave functions.
While the tensor networks offer high-accuracy results
in  gapped systems, they 
have difficulties,
for example, in metallic systems with a large entanglement beyond the so-called area law  that requires large tensor dimensions for the convergence.
Meanwhile, the mVMC method offers  relatively accurate wave functions based on the long-ranged resonating
valence bond states that are free from difficulties originating from large entanglement.

Although the efforts mentioned above have made substantial progress for ground-state properties,
the numerical simulation of finite-temperature or time-dependent properties remains challenging.
In a straightforward implementation, an accurate ensemble average over excited states is required for finite-temperature
or time-dependent simulation.
 However, rigorous ensemble averages 
are practically also formidable to carry out  because they require complete spectra of target systems.
Contrary to the straightforward implementation of  the full ensemble average,
the quantum transfer Monte Carlo method~\cite{JPSJ.55.3354} and  the subsequently proposed finite-temperature Lanczos method\cite{PhysRevB.49.5065}, and  Hams--De Raedt algorithm\cite{PhysRevE.62.4365} demonstrated that 
 calculation within small numbers of pure states instead of that in the full Hilbert space is numerically sufficient for
accurate estimates of finite-temperature physical quantities.

Indeed, 
it has been addressed  that a single random pure state chosen as the infinite-temperature initial sample 
 may exactly reproduce the infinite- as well as finite-temperature properties 
in the imaginary-time evolution in the thermodynamic limit 
without the necessity of taking the ensemble average.~\cite{JPSJ.55.3354} 
In Ref.~\citen{JPSJ.55.3354}, the idea starts from the fact that the  
finite-temperature properties of an observable that at least commutes
with the Hamiltonian are obtained exactly by
the imaginary-time evolution of only one pure state defined by $|\Phi\rangle \equiv \sum_i|\Psi_i\rangle$. 
Namely, this pure state can replace the ensemble average 
in any system with any size  if
the summation is taken over all the orthonormal eigenstates $\{|\Psi_i\rangle\}$ of
the Hamiltonian $\hat{H}$.
In the thermodynamic limit, physical properties calculated from $|\Phi\rangle$ and a single random initial state 
$|\Phi_r \rangle \equiv \sum_i r_{i} \vert\Psi_i \rangle$
with random coefficients $r_{i}$  converge to the same value irrespective of the commutativity of the 
physical variable with the Hamiltonian. This is for the following reason. For each infinitesimally small energy window between $E$ and $E+\delta E$, the summation of $|r_i|^2$ over the energy eigenstates $|\Psi_i\rangle$ with the energy $E_i$ that belongs to this energy window, namely $r_E$ defined by $r_E^2\delta E=\sum_{\{i|E\le E_i<E+\delta E\}} |r_i|^2$, converges to a unique value for each $E$ irrespective of the choice of random $r_i$ owing to the law of large numbers. This means that the averages calculated from $|\Phi \rangle$ and $|\Phi_r \rangle$ are the same within the microcanonical ensemble. The same is true for $\sum_{ij} r_i^*r_j A_{ij}$, where a matrix element of any physical quantity $A$ is denoted by $A_{ij}=\langle \Psi_i |A|\Psi_j\rangle$. After the energy integration, the canonical ensemble average can also be calculated only from $|\Phi_r \rangle$.

Recently, it was readdressed and proven that a single pure state gives us
thermodynamic quantities within errors well bounded by the inverse of the system size with an exponentially small prefactor
depending on the entropy of the system\cite{PhysRevLett.108.240401,PhysRevLett.111.010401,Sugiura_Shimizu_arXiv,PhysRevB.90.121110}.
Such pure states that
replace
the ensemble  are called 
 the thermal pure quantum (TPQ) states.
 The existence of the TPQ states corroborates the above already known idea and numerical procedures.
At finite temperatures, $\beta^{-1}=k_{\rm B}T$,
a Lanczos-type method\cite{PhysRevLett.108.240401} or  imaginary-time evolution initialized with a  small number of normalized random vectors
$\left|\Psi_0\right\rangle$ as $e^{-\frac{\beta}{2}\hat{H}}\left|\Psi_0\right\rangle$\cite{JPSJ.55.3354,PhysRevLett.111.010401},
where $\hat{H}$ is the Hamiltonian of the target system,  accurately simulates finite-temperature properties.
Later in this paper, we call the TPQ states without truncating the Hilbert space the full-space TPQ (FS-TPQ) states to distinguish them
from the TPQ states constructed in truncated Hilbert spaces.
The FS-TPQ states provide us with, as in the Lanczos method, the exact result within the statistical error
arising from the choice of the random initial states. An example for a small system size is shown in Appendix~A.

 However, beyond the system sizes tractable by the  previous FS-TPQ methods~\cite{JPSJ.55.3354,PhysRevB.49.5065,PhysRevE.62.4365}, we have to truncate the Hilbert space to construct the TPQ states.
In this paper, we propose a numerical scheme 
beyond the applicable range of the FS-TPQ states
by simulating the imaginary-time evolution of linear combinations of 
many-variable variational wave functions.
 The imaginary-time evolution is approximated by using 
the time-dependent variational principle (TDVP)\cite{mclachlan1964variational}, 
which is equivalent to the SR optimization  developed for the ground state\cite{arXiv.1507.00274}.
We call
the present method  the
finite-temperature variational Monte Carlo (FT-VMC) method.
The FT-VMC method based on the TDVP and many-variable variational wave functions is free from the negative sign problem
and  is applicable to a wide range of strongly correlated electron 
systems, as shown in the previous mVMC studies.~\cite{doi:10.1143/JPSJ.80.023704,
PhysRevB.83.205122,doi:10.1143/JPSJ.81.034701,PhysRevLett.108.177007,PhysRevB.90.115137,PhysRevB.89.195139,misawa2014superconductivity,doi:10.7566/JPSJ.83.093707,PhysRevB.92.035122,JPSJ.84.024720}

This paper is organized as follows.
In Sect.~2,
we explain details of the FT-VMC method.
In Sect.~3, we examine 
the accuracy and efficiency of the FT-VMC method 
by calculating the finite-temperature properties of the Hubbard model at half-filling,
which is a typical model Hamiltonian of strongly correlated electron systems,
in comparison with results of the FS-TPQ method.
We compare the results of the FT-VMC method with
the FS-TPQ method for the 16-site Hubbard ring
and the Hubbard model on a 4 by 4 square 
lattice with periodic boundary conditions.
We also examine the accuracy of the FT-VMC method
for larger system size by calculating the 24-site Hubbard ring.
Section 4 is devoted to a summary and discussion.

%
\section{Finite-Temperature Variational Monte Carlo Method}
%
\subsection{Expression of finite-temperature properties by pure states }
Thermodynamic quantities such as internal energy and specific heat
are in principle obtained by  taking ensemble averages.
However, there are pure states that can rigorously replace the ensemble in the thermodynamic limit, where the average of physical variables becomes the same. 
At the infinite temperature, the ensemble average of the system is proven to be replaced by the expectation value for a 
random vector~\cite{RIMS,PhysRevLett.99.160404}.

By lowering the temperature with the imaginary-time evolution, 
we can construct pure states.
It has indeed been shown that
the ensemble averages at finite temperatures are reasonably well reproduced
by taking averages over a small number of  pure states, as demonstrated by
the quantum transfer Monte Carlo method\cite{JPSJ.55.3354},
the finite-temperature Lanczos method\cite{PhysRevB.49.5065}
and
the Hams--De Raedt algorithm for finding
eigenvalues of huge matrices\cite{PhysRevE.62.4365},
and the microcanonical/canonical TPQ states\cite{PhysRevLett.108.240401,PhysRevLett.111.010401}.

Here, we summarize how to construct these pure 
states by following Refs.~\citen{JPSJ.55.3354,PhysRevB.49.5065,PhysRevE.62.4365} and \citen{PhysRevLett.111.010401}
for the formulation of the FT-VMC method detailed later.
We randomly pick up an orthonormal basis 
set ${\{\ket{i}\}}$ containing $L$ dimensions (components) out of a Hilbert space $\mathcal{H}$
and take random complex numbers $\{c_i\}$ that 
satisfy the normalization condition $\sum_i^L |c_i|^2 = 1$.
Then, an initial wave function $\ket{{\Psi}_0}$ is given by
 \begin{eqnarray}
\ket{\Psi_0}=\sum_{i}^L c_i \ket{i}.
 \end{eqnarray}

In addition, we define a vector  as
 \begin{eqnarray}
\ket{{\Psi}(\beta)}\equiv \exp(-\beta N_s\hat{h} / 2)\ket{{\Psi}_0},
 \end{eqnarray}
where $\hat{h}=\hat{H}/N_s$ is the Hamiltonian per site, $N_s$ is the number of sites, 
and $\beta$ is a constant identified as the inverse temperature below.
Since the state $\ket{{\Psi}_0}$ represents the ensemble of the system at the infinite temperature,
$\braket{\hat{A}}^{\mbox{\tiny ens}}_{\beta, N_s}\equiv \mbox{Tr}(e^{-N_s\beta \hat{h}}\hat{A})/\mbox{Tr}(e^{-N_s\beta \hat{h}})$ is equivalent to
the average over the random vectors $\left|{\Psi}_0\right\rangle$ as
\begin{eqnarray}
\braket{\hat{A}}^{\mbox{\tiny ens}}_{\beta, N_s}=
\overline{ \braket{\Psi(\beta)|\hat{A} |\Psi(\beta)}}/\overline{\braket{\Psi(\beta)|\Psi(\beta)}},
\end{eqnarray}
where $\overline{A}$ denotes a random average of a scalar $A$.
Sugiura and Shimizu~\cite{PhysRevLett.108.240401,PhysRevLett.111.010401} proved that the expectation value of physical variables $\hat{A}$ by $\ket{{\Psi}(\beta)}$ [$\braket{{\Psi}(\beta)|\hat{A}|{\Psi}(\beta)}/\braket{{\Psi}(\beta)|{\Psi}(\beta)}$] converges to $\braket{\hat{A}}^{\mbox{\tiny ens}}_{\beta, N_s}$ in the thermodynamic limit.
The pure state $\ket{\Psi(\beta)}$ is nothing but  the TPQ state.

It is easy to numerically operate a Hamiltonian to pure states and construct a TPQ state $\ket{{\Psi(\beta)}}$
when we keep wave functions $\ket{{\Psi}(\beta)}$ without truncating the full Hilbert space.
However, we will be faced with severe difficulties when
constructing TPQ states in truncated Hilbert spaces.
To construct TPQ states after truncating the Hilbert space, we employ
the TDVP with the help of an extended mVMC method introduced in the following sections.

\subsection{Time-dependent variational principle}
Here, we introduce the
TDVP\cite{mclachlan1964variational} for imaginary-time evolution in a truncated Hilbert space.
We start with the Schr\"odinger equation evolving in the imaginary time $\tau$ in the form
\begin{eqnarray}
\frac{d}{d\tau}\ket{{\Psi}(\tau)}=-\hat{H}\ket{{\Psi}(\tau)}.\label{iTDS}
\end{eqnarray}

If we take a trial wave function $\ket{\psi(\alpha(\tau))}$ parametrized by a set of $M$ variational parameters
$\alpha(\tau)=(\alpha_1(\tau),\alpha_2(\tau), \dots,\alpha_M (\tau))$, the imaginary-time evolution of the renormalized wave function
\begin{align}
\ket{\bar{\psi}(\alpha(\tau))}=\ket{\psi(\alpha(\tau))}/\sqrt{\braket{\psi(\alpha(\tau))|\psi(\alpha(\tau))}}
\end{align}
is given by rewriting Eq.~(\ref{iTDS}) as
\begin{align}
&\frac{d}{d\tau}\ket{\bar{\psi}(\alpha(\tau))}=\sum_{k}\dot{\alpha}_k\ket{\partial_{\alpha_k}\bar{\psi}(\alpha(\tau))}\nonumber\\
=&-(\hat{H}-\langle\hat{H}\rangle)\ket{\bar{\psi}(\alpha(\tau))}.\label{iTDS2}
\end{align}

The best approximation of this equation in the truncated Hilbert space
is obtained by  minimizing the difference between both sides of Eq.~(\ref{iTDS2}) defined as
\begin{align}
\left\|\sum_{k}\dot{\alpha}_k\ket{\partial_{\alpha_k}\bar{\psi}(\alpha(\tau))}+(\hat{H}-\langle\hat{H}\rangle)\ket{\bar{\psi}(\alpha(\tau))}\right\|,\label{iTDS3}
\end{align} 
where $\dot{\alpha}_k$ is the derivative of the $k$th variational parameter $\alpha_k$ with respect to $\tau$.
This principle is called the TDVP, and it is commonly applied to real-time evolution.

By projecting both sides of Eq.~(\ref{iTDS2}) by $\ket{\partial_{\alpha_m}\psi}$ to obtain the gradient vector  for minimizing Eq.~(\ref{iTDS3}),
we derive the following equations:
\begin{align} 
\sum_{k}\dot{\alpha}_k\braket{\partial_{\alpha_m}\bar{\psi}|\partial_{\alpha_k}\bar{\psi}}&=-\braket{\partial_{\alpha_m}\bar{\psi}|(\hat{H}-\langle\hat{H}\rangle)|\bar{\psi}},\\
\sum_{k}\dot{\alpha}_k\braket{\partial_{\alpha_k}\bar{\psi}|\partial_{\alpha_m}\bar{\psi}}&=-\braket{\bar{\psi}|(\hat{H}-\langle\hat{H}\rangle)|\partial_{\alpha_m}\bar{\psi}}, 
\end{align}
where we assume that the variational parameters $\alpha_k$ are real without loss of generality because we can treat real and imaginary part of parameters separately.
From the above set of equations, the following relation is derived:
\begin{align}
\sum_{k}\dot{\alpha}_k{\rm Re}\braket{\partial_{\alpha_k}\bar{\psi}|\partial_{\alpha_m}\bar{\psi}}=-{\rm Re}\braket{\bar{\psi}|(\hat{H}-\langle\hat{H}\rangle)|\partial_{\alpha_m}\bar{\psi}}.
\end{align}

By discretizing the derivative $\dot{\alpha}_k$ as $\Delta\alpha_k/\Delta \tau$,
we obtain the formula implemented in our numerical algorithm as
\begin{align}
\Delta \alpha_k=-\Delta \tau \sum_{m}S_{km}^{-1}g_{m},\label{iTDVP}
\end{align}

where
\begin{eqnarray}
S_{km}&\equiv&{\rm Re}\braket{\partial_{\alpha_k}\bar{\psi}|\partial_{\alpha_m}\bar{\psi}}\nonumber\\
&=&{\rm Re}\langle {\hat{O}_k \hat{O}_m}\rangle-{\rm Re}\langle {\hat{O}_k} \rangle{\rm Re}\langle{\hat{O}_m}\rangle
\end{eqnarray}
and
\begin{eqnarray}
g_{m}&\equiv&{\rm Re}\braket{\bar{\psi}|(\hat{H}-\langle\hat{H}\rangle)|\partial_{\alpha_m}\bar{\psi}}\nonumber\\
&=&{\rm Re}\langle {\hat{H} \hat{O}_m}\rangle-\langle {\hat{H}} \rangle{\rm Re}\langle{\hat{O}_m}\rangle.
\end{eqnarray}
Here, we define the operator $\hat{O}_k$ as
\begin{align}
\hat{O}_k=\sum_x\ket{x}\left(\frac{1}{\braket{x|\psi}}\frac{\partial}{\partial \alpha_k}\braket{x|\psi}\right)\bra{x},
\end{align}
where $\ket{x}$ is the real space configuration of electrons.
Here we note that the set of the real-space configuration $\{\ket{x}\}$ is orthogonal and complete.

The imaginary-time evolution given by Eq.~(\ref{iTDVP}) is nothing but the SR optimization
introduced by Sorella\cite{PhysRevB.64.024512}.
The SR method is a stable optimization method inspired by the imaginary-time evolution implemented by QMC sampling
and is used in the mVMC method to optimize the many-variable variational wave functions.
Even for the real-time evolution, the SR optimization is equivalent to the TDVP\cite{arXiv.1507.00274}.
To stabilize this optimization, we modify the diagonal elements in $S$ as
$S_{kk}\rightarrow S_{kk}(1+\epsilon_{\rm diag})$, and we disregard the elements of $S$ corresponding to $S_{kk}<\epsilon_{\rm wf}$, where $\epsilon_{\rm diag}, \epsilon_{\rm wf}\approx 1.0\times10^{-6}$\cite{JPSJ.77.114701}.

\subsection{Trial wave function}
In this paper, we span the truncated Hilbert space by the linear combination of
$N_{\rm pf}$ many-variable variational wave functions defined as
\begin{align}
|\psi\rangle=\sum^{N_{\rm pf}}_{n=1}{\mathcal P}^{n}_{\rm G} {\mathcal P}^{n}_{\rm J} |\phi^{n}_{\rm backflow}\rangle,\label{FT_VMC_WF}
\end{align}
where index $n$ runs up to $N_{\rm pf}$. For simplicity of notation, in this subsection, we drop the index $n$.
Here, we employ the generalized pairing wave function $|\phi_{\rm backflow}\rangle$ taking account of the backflow effect defined below [we define $\hat{c}^{\dag}_{i\sigma}$  ($\hat{c}_{i\sigma}$) as a creation (annihilation) operator of an electron with spin $\sigma$ on the $i$th site, and  $\hat{n}_{i\sigma}=\hat{c}^{\dag}_{i\sigma}\hat{c}_{i\sigma}$], and 
multiply by the Gutzwiller factor ${\mathcal P}_{\rm G}$ as well as the Jastrow factor ${\mathcal P}_{\rm J}$,
defined as
\begin{align}
{\mathcal P}_{\rm G}&=\exp\left(-g\sum_{i}\hat{n}_{i\uparrow}\hat{n}_{i\downarrow}\right),\\
{\mathcal P}_{\rm J}&=\exp\left(-\frac{1}{2}\sum_{i,j}v_{ij}\hat{n}_i \hat{n}_j\right).
\end{align}
Here the variational parameters in the Jastrow projection factor are set to depend only on the distance between the $i$th and $j$th sites.

Before defining the generalized pair-product wave function with backflow correlations, we give the pair-product wave function, defined as
\begin{align}
|\phi_{\rm pair}\rangle=\left(\sum^{N_s}_{i,j=1}f_{ij}\hat{c}^{\dag}_{i\uparrow}\hat{c}^{\dag}_{j\downarrow}\right)^{N_e/2}|0\rangle,
\end{align}
where $N_s$ is the number of sites and  $N_e$ is the number of electrons. 
To improve the pair-product wave function $|\phi_{\rm pair}\rangle$,
we take into account backflow correlations introduced
by Feynman and Cohen in their studies on liquid helium\cite{PhysRev.102.1189},
and by Tocchio  et al.
for Hubbard-like Hamiltonians\cite{PhysRevB.83.195138}.
Recently, some of the authors have introduced  backflow correlations into
the pair-product wave function $|\phi_{\rm pair}\rangle$\cite{arXiv.1507.00274}.

The pair-product wave function with backflow correlations is defined as
\begin{eqnarray}
\ket{\phi_{\rm backflow}} &=& \sum_{x} \ket{x} \braket{x | \phi_{\rm backflow}} \nonumber \\ 
&=& \sum_{x}\left( \frac{N_e}{2} \right)! {\rm Pf} \left[ X^b(x) \right]  \ket{x},
\end{eqnarray}
where ${\rm Pf} X^b(x)$ represents the Pfaffian of the skew-symmetric matrix $X^b(x)$. $X^b(x)$ is defined as
\begin{eqnarray}
 {X^b}_{nm}(x) = f^b_{r_n r_m}(x) -  f^b_{r_m  r_n}(x) .
\end{eqnarray}
The site of the $n$th ($m$th) electron is represented by $r_n$ ($r_m$). 
Here, we define the pair orbital with backflow correlations as
\begin{eqnarray}
&f_{r_n r_m}^b(x) = {\displaystyle \sum_{\mu,\nu=1}^{4} {\sum_{{\tau},{\tau}'}}} \eta^{\mu\nu}_{{\tau} {\tau}'} \Theta_{{r_n},{r_n}+\tau}^{\mu\uparrow}(x) \Theta_{r_m,r_m+\tau'}^{\nu\downarrow}(x) \nonumber \\
&\times f_{r_n+\tau, r_m+\tau'},
\end{eqnarray}
where 
\begin{eqnarray}
\Theta_{i,i+\tau}^{1\sigma}(x) &=& \delta_{i,i+\tau}, 
\\
\Theta_{i,i+\tau}^{2\sigma}(x) &=& \braket{\hat{D}_{i} \hat{H}_{i+\tau}}_x, 
\\
\Theta_{i,i+\tau}^{3\sigma}(x) &=&  \braket{\hat{n}_{i\sigma} \hat{h}_{i,-\sigma} \hat{n}_{i+\tau,-\sigma} \hat{h}_{i+\tau\sigma}}_x,
\end{eqnarray}
and
\begin{eqnarray}
\Theta_{i,i+\tau}^{4\sigma}(x) = \braket{\hat{D}_{i}\hat{n}_{i+\tau,-\sigma} \hat{h}_{i+\tau\sigma}+\hat{n}_{i\sigma} \hat{h}_{i,-\sigma} \hat{H}_{i+\tau}}_x.
\end{eqnarray} 
In the above representation, $\eta^{\mu,\nu}_{\tau,\tau'}$ are variational parameters
 and subscripts $\tau$ ($\tau'$) represent the relative distance from $r_n$ ($r_m$).
These subscripts run  up to $\ell$th-neighbor sites, where $\ell$ controls the range of backflow correlations.
We set $\ell=2$ ($\ell=1$) for one-dimensional (two-dimensional) systems.
In other words, we take into account the nearest- and next-nearest-neighbor correlation in the one-dimensional case and only the nearest-neighbor correlation in the two-dimensional case.
We assume that $\eta^{\mu\nu}_{{\tau} {\tau}'}$ depends on distance, i.e., $\eta^{\mu\nu}_{{\tau} {\tau}'}=\eta^{\mu\nu}_{-{\tau} {\tau}'}=\eta^{\mu\nu}_{{\tau} -{\tau}'}=\eta^{\mu\nu}_{-{\tau} -{\tau}'}$.
In addition, $\delta_{ij}$ is the Kronecker delta, $\hat{D}_i=\hat{n}_{i\uparrow}\hat{n}_{i\downarrow}, \hat{H}_i=(1-\hat{n}_{i\uparrow})(1-\hat{n}_{i\downarrow})$, $\hat{h}_{i\sigma}=1-\hat{n}_{i\sigma}$,
and $\langle \hat{A}\rangle_x=\langle x|\hat{A}|x\rangle/\langle x|x\rangle$.
We assume that
$\eta^{1,1}_{\tau,\tau'}(x)=1$ if
$\Theta_{i,i+\tau}^{2\sigma}(x)=0$ for any $\tau$ and $\sigma$.

We take the variational parameters of one-body parts, $f_{ij}$, as complex numbers and Gutzwiller--Jastrow
projection factors and backflow correlations, $g$, $v_{ij}$, and $\eta^{\mu,\nu}_{\tau,\tau'}$ as real numbers.
The total number of variational parameters is $N_{\rm tot}=N_{\rm pf}(2N_{s}^2+N_{\rm G\mathchar`-proj}+N_{\rm J\mathchar`-proj}+N_{\rm bf})$, where
${N_{\rm pf}}$, ${N_{\rm G\mathchar`-proj}}$, ${N_{\rm J\mathchar`-proj}}$, and $N_{\rm bf}$
are the numbers of Pfaffians, Gutzwiller projection factors, Jastrow projection factors, and backflow parameters, respectively.
The number of  variational parameters employed in the present paper is summarized in Table \ref{num_vp}.
We note that, for $N_{\rm pf}=10$, the number of variational parameters is about $5\times10^3$ for a 16-site system, which
is much smaller than the total number of dimensions of the Hilbert space  $L\sim1.6\times10^{8}$.
\begin{table}[ht]
\begin{tabular}{lc|lc}
\hline
\hline
1D Hubbard & & 2D Hubbard &\\
\hline
$N_{\rm s}$ & 16 & $N_{\rm s}$ & 16 \\
\hline
$N_{\rm G\mathchar`-proj}$ & 1 & $N_{\rm G\mathchar`-proj}$ & 1 \\
$N_{\rm J\mathchar`-proj}$ & 8 & $N_{\rm J\mathchar`-proj}$ & 5 \\
$N_{\rm bf}$ & 28 & $N_{\rm bf}$ & 10 \\
\hline
$N_{\rm tot}$
& $549N_{\rm pf}$&
$N_{\rm tot}$
& $528N_{\rm pf}$\\
\hline
\hline
\end{tabular}
\caption{Number of variational parameters for one-dimensional (1D) and
two-dimensional (2D) Hubbard models.}
\label{num_vp}
\end{table}

\subsection{Whole procedure of proposed method}
Here, we summarize the whole procedure of the proposed method.
\begin{description}
\item[First step]{Choice of initial states}\\
In
the original method involving FS-TPQ states,
we take random vectors as initial states.  
However, it is not practically possible to prepare
random vectors by using the variational wave functions $\left|\psi\right\rangle$ defined in Eq.~(\ref{FT_VMC_WF}),
which span a truncated Hilbert space.
To mimic the random vectors within the truncated Hilbert space, we employ
a linear combination of  Pfaffian wave functions $\ket{\phi^n}$ with random elements $f_{ij}^{n}$ as the initial wave function, i.e.,
\begin{eqnarray}
\ket{\psi(\alpha(\tau=0))}=\sum^{N_{\rm pf}}_{{n}=1}\ket{\phi^{{n}}},
\end{eqnarray}
where
\begin{eqnarray}
\ket{\phi^n}=\left(\sum^{N_s}_{i,j=1}f^{n}_{ij}\hat{c}^{\dag}_{i\uparrow}\hat{c}^{\dag}_{j\downarrow}\right)^{\frac{N_e}{2}}\ket{0}.
\end{eqnarray}
Here, we choose $f^{n}_{ij}=f^1_{ij}+\epsilon^{n}_{ij}$ $( {\rm with} \;\;|\epsilon^{n}_{ij}/f^1_{ij}|\simeq 0.01\;\;{\rm for}\; n \neq 1)$
to make finite overlaps among the Pfaffian wave functions $\ket{\phi^{n}}$.

\item[Second step]{Construction of TPQ states}\\
Next, we construct a TPQ state by increasing the inverse temperature by $\Delta \tau$ as
\begin{align}
&\ket{\bar{\psi}(\alpha(\tau+\Delta\tau))}\nonumber\\
=&\ket{\bar{\psi}(\alpha(\tau))}-(\hat{H}-\langle\hat{H}\rangle)\Delta \tau\ket{\bar{\psi}(\alpha(\tau))}+O({(\Delta\tau)}^2).
\label{timeevolve}
\end{align}
In this process, we employ the TDVP formulated in Eq.~(\ref{iTDVP}).

\item[Third step]{Calculation of observables}\\
The approximate FS-TPQ state
$\ket{\bar{\psi}(\alpha(\tau+\Delta\tau))}$
obtained in the above step replaces the ensemble average at $k_{\rm B}T=\frac{1}{2}(\tau+\Delta\tau)^{-1}$.
Therefore,
the physical quantities at $k_{\rm B}T=\frac{1}{2}(\tau+\Delta\tau)^{-1}$ are given by
the expectation values calculated only with $\ket{\bar{\psi}(\alpha(\tau+\Delta\tau))}$.
\end{description}
The second and third steps are repeated
until $\beta=2\tau$ reaches the required inverse temperature.

\begin{figure}[ht]
\begin{center}
\includegraphics[width=7cm]{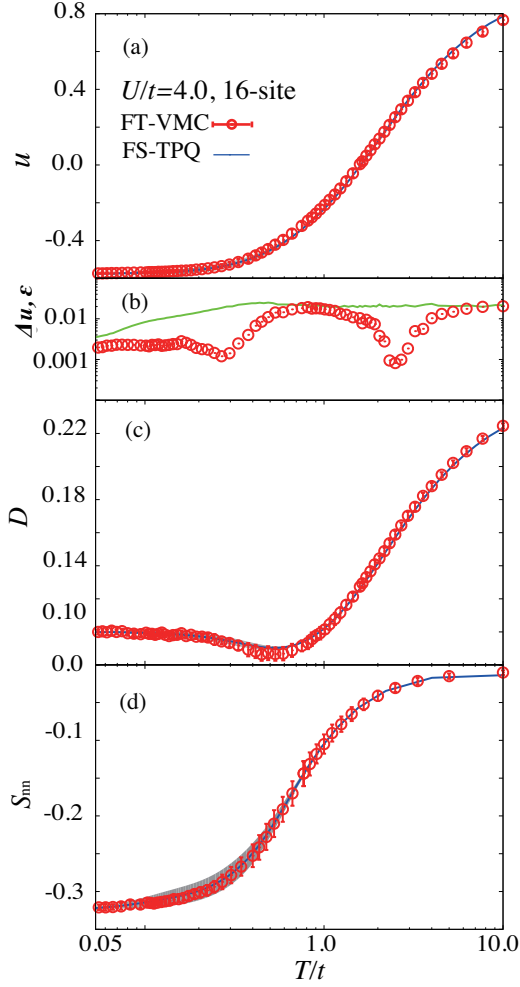}
\end{center}
\caption{ \label{fig:1dU4}(color online) 
(a)~Temperature dependence of internal energy 
per site $u$ calculated by the FT-VMC method
for a 16-site half-filled Hubbard ring with $N_{\uparrow}=N_{\downarrow}=8$ and $U/t=4.0$.
The FT-VMC result is shown by the (red) circles 
in comparison with the results of the FS-TPQ method shown by the (blue) curve.
(b)~The (Green) solid curve represents the estimated errors $\epsilon$ in the FT-VMC (FS-TPQ) simulation
and (red) circles show absolute values of differences $\Delta u$ between the 
results of the FT-VMC and FS-TPQ methods. 
(c)~Temperature dependence of double occupancy $D$ per site calculated by the FT-VMC  and FS-TPQ methods.
The results of the FT-VMC and FS-TPQ methods are shown by the (red) circles and  (blue) curve, respectively.
(d)~Nearest-neighbor spin correlation $S_{\rm nn}$
calculated by the FT-VMC and FS-TPQ methods.
The results of the FT-VMC and FS-TPQ methods are shown by the (red) circles and  (blue) curve, respectively.
The shaded regions in (a), (c), and (d) represent the error bars of
$u$, $D$, and $S_{\rm nn}$  obtained from the FS-TPQ method, respectively.}
\end{figure}

%
\section{Benchmark Results}
%
We compare the results  obtained by the FT-VMC method and the FS-TPQ methods.
The accuracy of the FS-TPQ method is examined in Appendix~B.
In this paper, we calculate the finite-temperature properties of
the standard Hubbard model defined by
\begin{eqnarray}
\hat{H}=-t\sum_{\langle i,j\rangle,\sigma}(\hat{c}^{\dag}_{i\sigma}\hat{c}_{j\sigma}+\mbox{h.c.})+U\sum_{i}\hat{n}_{i\uparrow}\hat{n}_{i\downarrow},
\end{eqnarray}
where $t$ is the hopping amplitude between the nearest-neighbor sites $\langle i,j\rangle$, 
$\hat{c}^{\dag}_{i\sigma}$ ($\hat{c}_{i\sigma}$) is the creation (annihilation) operator of an electron with spin $\sigma$ on the $i$th site,
and $U$ is the on-site Coulomb repulsion with $\hat{n}_{i\sigma}=\hat{c}^{\dag}_{i\sigma}\hat{c}_{i\sigma}$.

The Hamiltonian of the Hubbard model conserves the numbers of spin-up 
electrons $N_{\uparrow}$ and spin-down electrons $N_{\downarrow}$. 
Here we impose  $N_{\uparrow}=N_{\downarrow}$  to reduce computational costs. 
In finite-size systems,
the results obtained under this condition are not the 
same as those obtained by the usual canonical ensemble average,
where the average is taken over the states with $N_{\uparrow}\neq N_{\downarrow}$ in addition to the states with $N_{\uparrow}= N_{\downarrow}$.
Both results agree with each other in the thermodynamic limit.

In this paper, we estimated errors as the standard deviation of each calculation.
Hereafter, we take $k_{\rm B}=1.0$.

\begin{figure}[h!]
\begin{center}
\includegraphics[width=7cm]{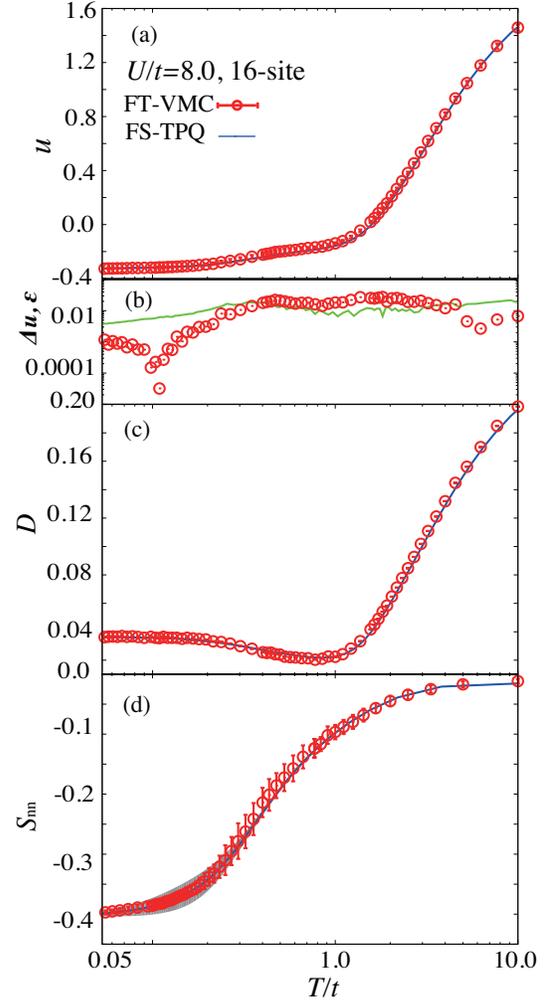}
\end{center}
\caption{\label{fig:1dU8}(color online) 
(a)~Temperature dependence of internal energy 
per site $u$ calculated by the FT-VMC method
for a 16-site half-filled Hubbard ring with $N_{\uparrow}=N_{\downarrow}=8$ and $U/t=8.0$.
The FT-VMC result is shown by the (red) circles 
in comparison with the results of the FS-TPQ method shown by the (blue) curve.
(b)~The (Green) solid curve represents the estimated errors $\epsilon$ in the FT-VMC (FS-TPQ) simulation
and (red) circles show absolute values of differences $\Delta u$ between the 
results of the FT-VMC and FS-TPQ methods. 
(c)~Temperature dependence of double occupancy $D$ per site calculated by the FT-VMC  and FS-TPQ methods.
The results of the FT-VMC and FS-TPQ methods are shown by the (red) circles and  (blue) curve, respectively.
(d)~Nearest-neighbor spin correlation $S_{\rm nn}$
calculated by the FT-VMC and FS-TPQ methods.
The results of the FT-VMC and FS-TPQ methods are shown by the (red) circles and  (blue) curve, respectively.
The shaded regions in (a), (c), and (d) represent the error bars of
$u$, $D$, and $S_{\rm nn}$  obtained from the FS-TPQ method, respectively.}
\end{figure}

\subsection{One-dimensional Hubbard model}

First, we show the temperature dependence of the
physical properties of a 16-site Hubbard ring at 
half-filling ($N_{\uparrow}=N_{\downarrow}=8$ and $U/t=4.0$).
In Fig.~\ref{fig:1dU4}(a),
we compare the internal energy  per site $u=\langle\hat{H}\rangle/N_s$, with $N_s$ being the number of sites of the total system, obtained by the FT-VMC and FS-TPQ methods. 
We perform random averaging over 10 initial random vectors
starting from independent random states 
at $T/t = +\infty$ and we take a linear combination of 10 Pfaffians ($N_{\rm pf}=10$) as
the variational wave functions. We take $\Delta \tau = 0.025$.

As shown below, in the whole temperature range of the calculation,
the temperature dependence of the energy obtained by the FT-VMC method is  in
good agreement with that of the FS-TPQ method.
This result demonstrates that our variational wave function
can describe the FS-TPQ states with a high accuracy.

To examine the accuracy of the FT-VMC method, 
we compare the absolute values of 
differences
between the expectation values obtained from the FT-VMC  and FS-TPQ methods,
$\Delta u$,
which is defined as
\begin{align}
\Delta u=|u_{\rm FT\mathchar`-VMC}-u_{\rm FS\mathchar`-TPQ}|.
\end{align}
We also define the error of the internal energy $\epsilon$ by taking into account
the propagation of errors as
\begin{align}
\epsilon =\sqrt{\epsilon_{\rm FT\mathchar`-VMC}^2+\epsilon_{\rm FS\mathchar`-TPQ}^2},
\end{align}  
where $\epsilon_{\rm FT\mathchar`-VMC}$ ($\epsilon_{\rm FS\mathchar`-TPQ}$) is the random averaging error of the FT-VMC (FS-TPQ) method.
The FT-VMC result for $u$ is in good agreement with that for the FS-TPQ states
within the error bars as can be seen in Fig.~\ref{fig:1dU4}(a).
As shown in Fig.~\ref{fig:1dU4}(b), the difference in energy $\Delta u$ is lower than $\epsilon$ in the whole temperature range.

Next, we also compare the temperature dependence of 
the double occupancy per site $D$, which is defined as
\begin{align}
D=\sum_{i}\langle \hat{n}_{i\uparrow}\hat{n}_{i\downarrow} \rangle/N_{s}.
\end{align}
As shown in Fig.~\ref{fig:1dU4}(c),
the results of the FT-VMC method agree with those of the FS-TPQ method
within the error bars.

We also calculate the nearest-neighbor spin correlation, which is 
defined as

\begin{eqnarray}
S_{\rm nn}=\langle\vec{S}_{i}\cdot \vec{S}_{i+1}\rangle,
\end{eqnarray}
where $\vec{S}_i=(\hat{S}^x_i,\hat{S}^y_i,\hat{S}^z_i)^{T}$,
$\hat{S}^{\gamma}=1/2\sum_{\sigma,\sigma^{\prime}}
\hat{c}^{\dagger}_{i,\sigma}\sigma^{\gamma}_{\sigma,\sigma^{\prime}}\hat{c}_{i,\sigma^{\prime}}$
($\gamma=x,y,z$),
and $(\sigma^{x},\sigma^{y},\sigma^{z})$
are the Pauli matrices.
As shown in Fig.~\ref{fig:1dU4}(c),
we also confirm that the results obtained by the FT-VMC
method are  in good agreement with  those of the FS-TPQ method within the error bars.

To examine the range of applicability of our method,
we have perform the calculation in the strong-coupling region, i.e., $U/t=8.0$.
As shown in Figs.~\ref{fig:1dU8}(a)-\ref{fig:1dU8}(d), even in the strong-coupling region, 
 the results of the FT-VMC method are in
good agreement with those of the FS-TPQ method.

Although the results of  the FT-VMC method slightly deviate from those of
the FS-TPQ method  at intermediate temperatures, $T/t\sim 1.0$,
$\Delta u$ and $\epsilon$ are close to each other as shown in Fig.~\ref{fig:1dU8}(b).
As we discuss later in Sect.~\ref{conv},
this deviation may be resolved by increasing 
the number of  linear combinations of Pfaffian wave functions.

As shown in Figs.~\ref{fig:1dU8}(c) and 2(d),
the double occupancy $D$ and the nearest-neighbor spin correlation $S_{\rm nn}$
calculated by the FT-VMC method also agree with the results of the FS-TPQ method within their error bars.
In common with the intermediate- and strong-coupling regions ($U/t=4.0$ and $U/t=8.0$, respectively),
the minima in the temperature dependence of the double occupancy $D$ are well reproduced in the FT-VMC method,
which signal the development of antiferromagnetic spin correlations\cite{gorelik2012universal} as shown in
Fig.~\ref{fig:1dU8}(d).

\begin{figure}[h!]
\begin{center}
\includegraphics[width=8cm]{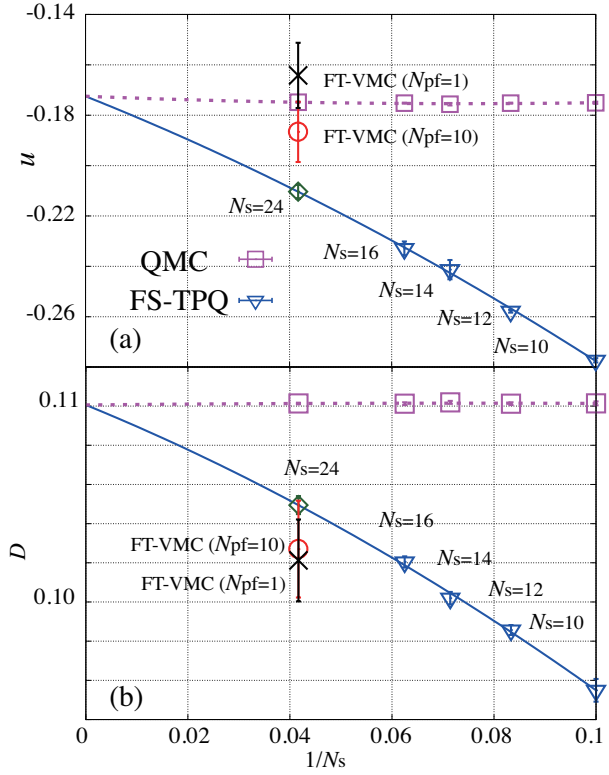}
\end{center}
\caption{\label{fig:intp}(color online)
Fittings of size dependences
for (a)~internal energy per site $u$ and (b) double occupancy $D$ calculated from the FS-TPQ [(blue) triangles]
and QMC [(purple) squares] methods
for half-filled Hubbard rings with $U/t=4.0$ at $T/t=1.0$.
To evaluate the expected FS-TPQ results at any given system size, first, we perform the size extrapolation of the QMC results and estimate $u$ and $D$ in the thermodynamic limit.
Then, we interpolate the estimated values in the thermodynamic limit and the available FS-TPQ results to estimate the exact results
at any given system size in the canonical ensemble.
The FT-VMC results of a 24-site Hubbard ring are shown by the (red) circles for $N_{\rm pf}=10$ and (black) crosses  for $N_{\rm pf}=1$
in comparison with the exact values estimated from the FS-TPQ and QMC methods shown by the (dark-green) diamonds.
The (blue) curves and (purple) dotted lines in (a) and (b) represent the interpolation and extrapolation curves of the FS-TPQ and QMC methods, respectively.
}
\end{figure}

In addition, we calculate a 24-site Hubbard ring at half-filling ($N_{\uparrow} =N_{\downarrow} =12$, $U/t=4.0$, $N_{\rm pf}=10$)
to investigate the accuracy of the FT-VMC method for  larger systems.
To examine the accuracy of our FT-VMC results with a fixed number of electrons, we need to compare our results with
the exact results obtained by using the canonical ensemble.
However, the QMC method at finite temperatures employs the grand canonical ensemble, which gives different expectation values
from those obtained by using the canonical ensemble for finite-size systems.
Consequently, it is not easy to compare the results of the FT-VMC and QMC methods.
Hence, we interpolate the results of the FS-TPQ method and those in the thermodynamic limit
estimated by using the grand canonical QMC method after size extrapolation by assuming the size scaling of a physical quantity
$\langle \hat{A}\rangle_{N_s} = \langle \hat{A}\rangle_{+\infty}+a/N_s+b/N_s^2+\mathcal{O}(1/N_s^3)$, where $\langle \hat{A}\rangle_{N_s}$ and $\langle \hat{A} \rangle_{+\infty}$ represent the expected values of $\langle \hat{A} \rangle$ for $N_s$ sites and those in the thermodynamic limit, respectively. 
By interpolation with a fitting, we can estimate exact results for 24 sites at a given temperature.
We use the FS-TPQ results for 10, 12, 14, and 16-site Hubbard rings for the interpolation
with the least-squares fitting as shown in Fig.~\ref{fig:intp}.
Here, note that both the QMC and FS-TPQ methods should converge to the same values in the  thermodynamic limit.
We show the FT-VMC results for $N_{\rm pf}=1$ and 10 for $N_s=24$. The results indicate that the agreement with the estimated exact values is improved by increasing $N_{\rm pf}$ as expected.

\begin{figure}[h!]
\begin{center}
\includegraphics[width=7cm]{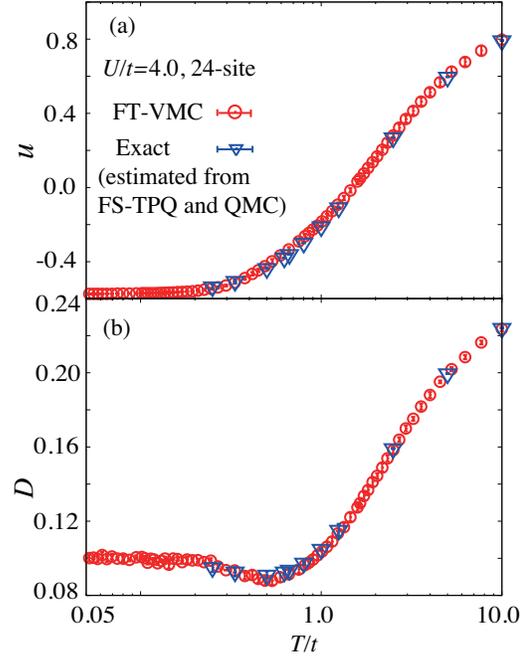}
\end{center}
\caption{\label{fig:1d24s}(color online)
Temperature dependences of (a) internal energy per site $u$ and (b) double occupancy $D$ per site, calculated from the FT-VMC
method for a $24$-site half-filled Hubbard ring with $N_{\uparrow}=N_{\downarrow}=12$ and $U/t=4.0$.
The FT-VMC results are shown by the (red)  circles
in comparison with the
exact
results estimated from the QMC and FS-TPQ methods shown by the (blue) triangles.
}
\end{figure}

In  Fig.~\ref{fig:1d24s}, we show the temperature dependences of 
the energy per site $u$ and double occupancy $D$. 
The FT-VMC results are shown for $N_{\rm pf}=10$, where the exact results are well reproduced.
In Fig.~\ref{fig:1d24s}, the error bars shown are those arising from the interpolation.

%

\begin{figure}[h!]
\begin{center}
\includegraphics[width=7cm]{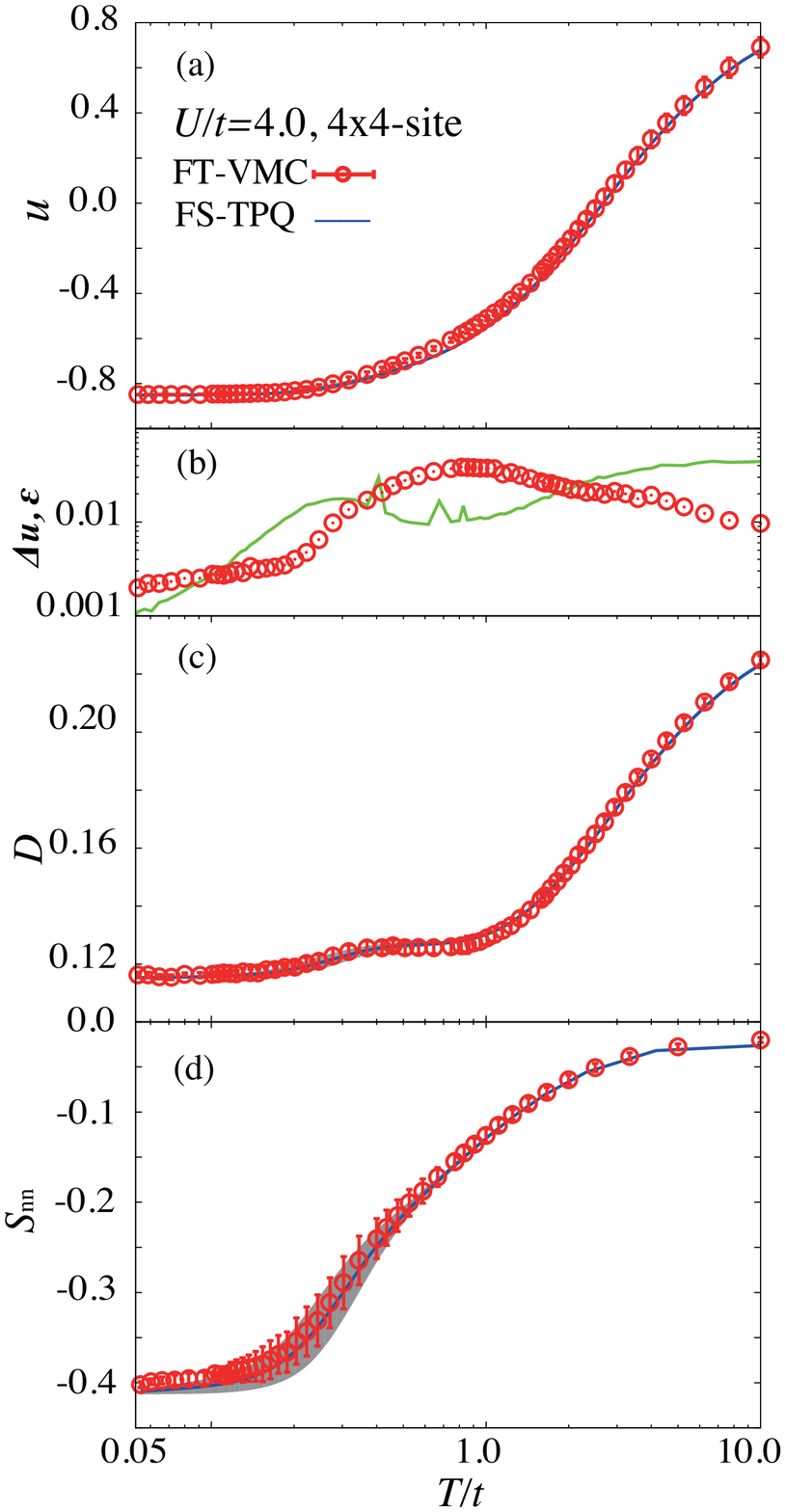}
\end{center}
\caption{\label{fig:2dU4}(color online)
(a)~Temperature dependence of internal energy 
per site $u$ calculated by the FT-VMC method
for a $4\times4$-site half-filled Hubbard model with $N_{\uparrow}=N_{\downarrow}=8$ and $U/t=4.0$.
The FT-VMC result is shown by the (red) circles 
in comparison with the results of the FS-TPQ method shown by the (blue) curve.
(b)~The (Green) solid curve represents the estimated errors $\epsilon$ in the FT-VMC (FS-TPQ) simulation
and (red) circles show absolute values of differences $\Delta u$ between the 
results of the FT-VMC and FS-TPQ methods. 
(c)~Temperature dependence of double occupancy $D$ per site calculated by the FT-VMC  and FS-TPQ methods.
The results of the FT-VMC and FS-TPQ methods are shown by the (red) circles and  (blue) curve, respectively.
(d)~Nearest-neighbor spin correlation $S_{\rm nn}$
calculated by the FT-VMC and FS-TPQ methods.
The results of the FT-VMC and FS-TPQ methods are shown by the (red) circles and  (blue) curve, respectively.
The shaded regions in (a), (c), and (d) represent the error bars of
$u$, $D$, and $S_{\rm nn}$  obtained from the FS-TPQ method, respectively.
}
\end{figure}

\subsection{Two-dimensional Hubbard model}
%

We next compare the results obtained by the FT-VMC 
and FS-TPQ methods
for the $4\times 4$-site  square-lattice Hubbard model at half-filling 
for $U/t = 4.0$ as shown in Fig.~\ref{fig:2dU4}. The computational conditions are the same as those for
the one-dimensional case.

We again see good agreement of the temperature dependence of the
energy between the FS-TPQ and FT-VMC methods in a wide range of temperatures.
However, in the intermediate temperature region ($T/t\sim 1.0$), 
$\Delta u$ is slightly larger than $\epsilon$. 
Except for the slight deviation of energy, 
we confirm that the temperature dependences of 
the double occupancy $D$ and the nearest-neighbor spin correlation $S_{\rm nn}$ have good consistency 
with those of the FS-TPQ method within the error bars.
This agreement of the physical properties guarantees that 
our FT-VMC method successfully describes the FS-TPQ states in two dimensions.

Here we note that the nonmonotonic temperature dependence of $\partial D/\partial T$
 shown in Fig.~\ref{fig:2dU4}(c) is also well reproduced.
Such nonmonotonic behavior is hardly expected in the perturbative and mean-field approaches,
which may prove the accuracy of the FT-VMC method.

\begin{figure}[h!]
\begin{center}
\includegraphics[width=7cm]{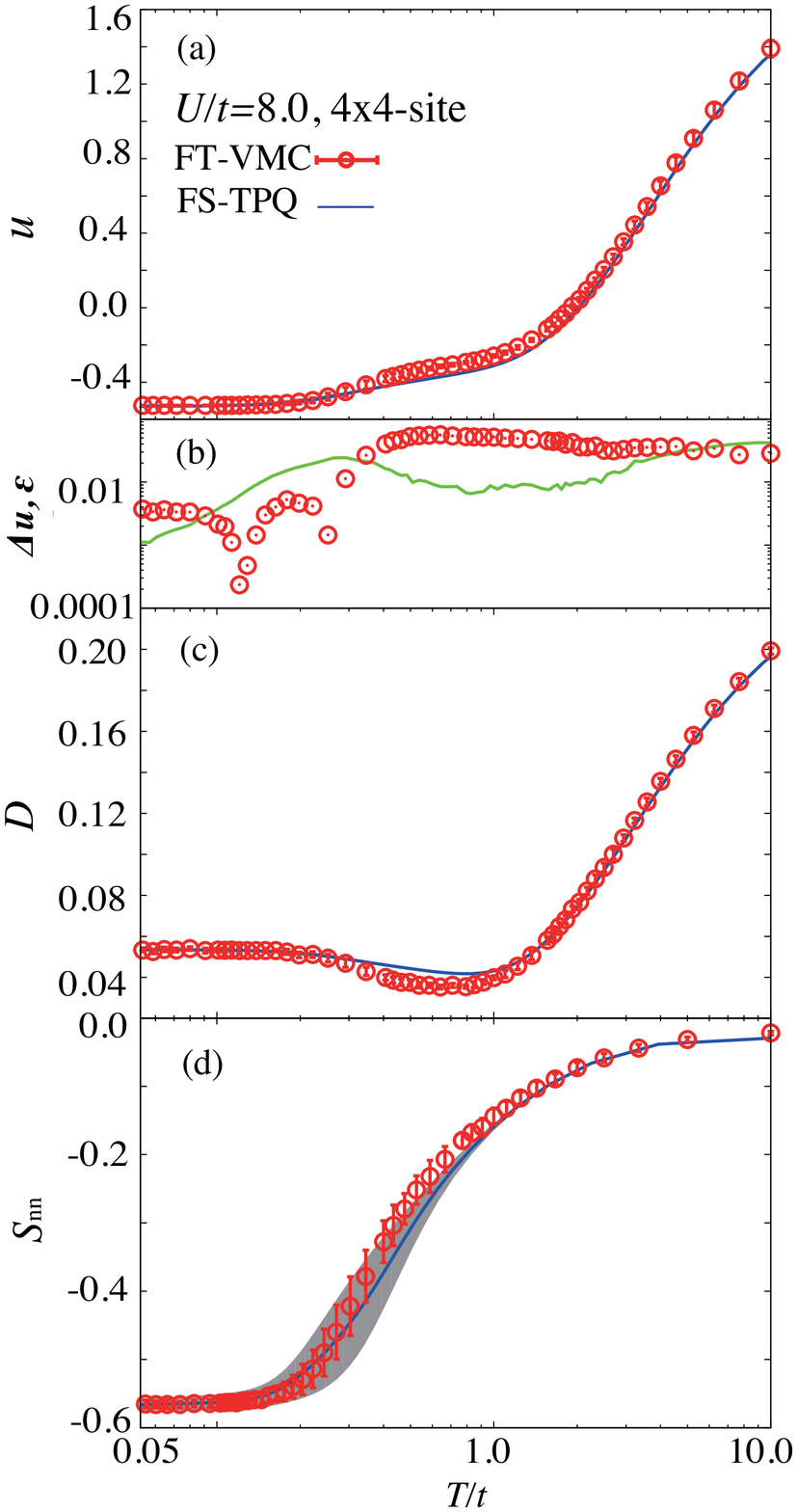}
\end{center}
\caption{\label{fig:2dU8}(color online)
(a)~Temperature dependence of internal energy 
per site $u$ calculated by the FT-VMC method
for a $4\times4$-site half-filled Hubbard model with $N_{\uparrow}=N_{\downarrow}=8$ and $U/t=8.0$.
The FT-VMC result is shown by the (red) circles 
in comparison with the results of the FS-TPQ method shown by the (blue) curve.
(b)~The (Green) solid curve represents the estimated errors $\epsilon$ in the FT-VMC (FS-TPQ) simulation
and (red) circles show absolute values of differences $\Delta u$ between the 
results of the FT-VMC and FS-TPQ methods. 
(c)~Temperature dependence of double occupancy $D$ per site calculated by the FT-VMC  and FS-TPQ methods.
The results of the FT-VMC and FS-TPQ methods are shown by the (red) circles and  (blue) curve, respectively.
(d)~Nearest-neighbor spin correlation $S_{\rm nn}$
calculated by the FT-VMC and FS-TPQ methods.
The results of the FT-VMC and FS-TPQ methods are shown by the (red) circles and  (blue) curve, respectively.
The shaded regions in (a), (c), and (d) represent the error bars of
$u$, $D$, and $S_{\rm nn}$  obtained from the FS-TPQ method, respectively.}
\end{figure}

We also perform the same calculations 
for the strong-coupling  region ($U/t=8.0$) 
as shown in Fig.~\ref{fig:2dU8}.
The qualitative behaviors of the physical properties are
well reproduced by the FT-VMC method.
However, within the linear combination of the trial wave functions comprising up to 10 Pfaffians, 
we find that $\Delta u$ is
larger than $\epsilon$ in  a wide range of temperatures.
We also find that the temperature dependences of the 
double occupancy $D$ and nearest-neighbor spin correlation $S_{\rm nn}$ deviate
from those obtained by the FS-TPQ method even though the qualitative trends of the temperature dependences
of $D$ and $S_{\rm nn}$ are captured in the FT-VMC results. 
This deviation indicates that a larger number of Pfaffian wave functions
is necessary in the strong-coupling region
for two-dimensional systems.

\subsection{Convergence of the results by increasing the number of Pfaffian wave functions}
\label{conv}
\begin{figure}[h!]
\begin{center}
\includegraphics[width=8cm]{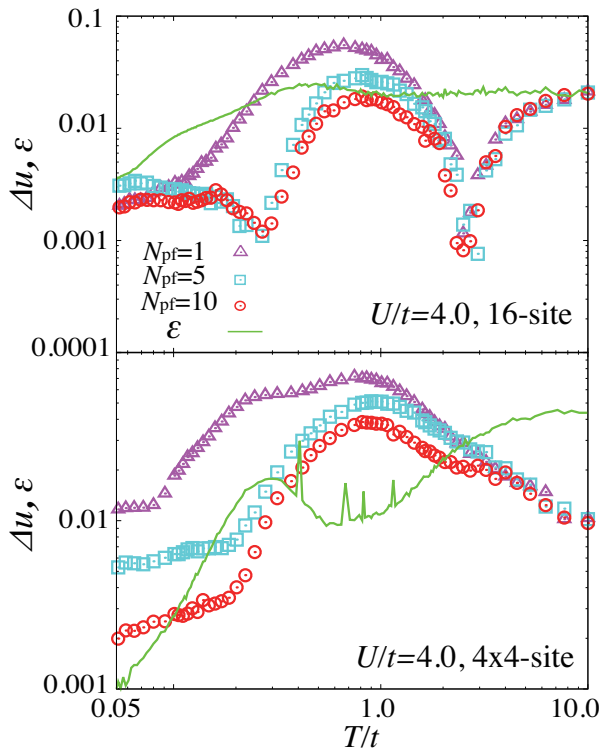}
\end{center}
\caption{\label{fig:diff}(color online)
(a)~$\Delta u$ and $\epsilon$ calculated from the FT-VMC method
for a $16$-site half-filled Hubbard model with $N_{\uparrow}=N_{\downarrow}=8$ and $U/t=4.0$.
The (green) solid line represents $\epsilon$ for $N_{\rm pf}=10$
and (purple) triangles,  (light-blue) squares, and   (red) circles 
respectively show absolute values of the differences $\Delta u$ for $N_{\rm pf}=1, 5$ and 
10 between the result of the FT-VMC and FS-TPQ methods.
(b)~$\Delta u$ and $\epsilon$ calculated from the FT-VMC method
for $4\times 4$-site half-filled Hubbard model with $N_{\uparrow}=N_{\downarrow}=8$ and $U/t=4.0$.
The (green) solid line represents $\epsilon$ for $N_{\rm pf}=10$
and  (purple) triangles,  (light-blue) squares, and   (red) circles 
respectively show absolute values of the differences $\Delta u$ for $N_{\rm pf}=1, 5$ and 
10 between the results of the FT-VMC and FS-TPQ methods.
}
\end{figure}

Here, we discuss the convergence of the results 
when we increase $N_{\rm pf}$ 
(the number of Pfaffian wave functions involved in the trial wave function).
To examine the convergence, the $N_{\rm pf}$ dependence of $\Delta u$ (the error of the energy per site $u$)
is shown in Fig.~\ref{fig:diff}.
Here,
we systematically change the number of Pfaffians up to $N_{\rm pf}=10$ ($N_{\rm pf}=1,5,10$).
We show $\Delta u$
and
$\epsilon$ (in the case of $N_{\rm pf}=10$) for the
16-site Hubbard ring with $U/t=4.0$ at half-filling
[the same as for the temperature dependences of $\Delta u$ and $\epsilon$ shown in Fig.~\ref{fig:1dU4}(b)] and the 4$\times$4-site Hubbard model with $U/t=4.0$ at half-filling
[the same as for the temperature dependences of $\Delta u$ and $\epsilon$ shown in Fig.~\ref{fig:2dU4}(b)].
 As expected, the accuracy of the FT-VMC method 
deteriorates in the intermediate-temperature region ($T/t\sim 1.0$),
namely, $\Delta u$ calculated by the FT-VMC method markedly deviates from that calculated by
the FS-TPQ method when we take $N_{\rm pf}=1$.
However, this deviation can be reduced by increasing $N_{\rm pf}$. 
We confirm that $\Delta u$ becomes smaller
as
$N_{\rm pf}$ is increased, and 
it converges within $\epsilon$, as is evident in Fig.~\ref{fig:diff}.

This result shows that we can systematically improve the accuracy
of the FT-VMC method by increasing the number of  
Pfaffian wave functions.
As we have shown above, in the strong-coupling region of two-dimensional systems,
we find that $N_{\rm pf}=10$ is not  sufficient for the convergence of results  at intermediate temperatures.
Because it requires an increased numerical cost and it is difficult to perform at this stage,
calculations with larger linear combinations of  Pfaffian wave functions ($N_{\rm pf}>10$) 
are left for future studies.
 Since the mVMC method is size consistent, it is expected that the number of Pfaffian states needed to reproduce the finite-temperature properties does not increase exponentially with the system size. A key issue be improved is to find a way of approximating the initial random states more correctly by using a linear combination of a moderate number of optimally overlapped Pfaffians.

%
\section{Summary and Outlook}
%
 We have developed a finite-temperature variational Monte Carlo (FT-VMC) method for strongly correlated electron systems and have shown benchmark results for the Hubbard models.
The method is based on the path integral in the imaginary-time direction. The initial state given by a small number of random Pfaffians well approximates the infinite-temperature starting point. The imaginary-time evolution combined with an optimal truncation of the Hilbert space that expands the variational wave function allows the treatment of finite-temperature properties at large system sizes beyond the limit tractable by the quantum transfer matrix method and the finite-temperature Lanczos method. 
In our method, trial wave functions that are optimized are regarded
as thermal pure quantum (TPQ) states at each temperature.
To obtain approximate TPQ states
in a truncated Hilbert space, 
we optimize variational wave functions by using the time-dependent variational principle.
In the benchmark calculations, we found that the FT-VMC results for
the temperature dependences of internal energy, double occupancy, and spin correlations in both one and two dimensions 
well reproduce the full-space TPQ (FS-TPQ) results, which are basically expected to be exact.
These results show that our trial wave function offers a highly accurate and efficient description of TPQ states in the whole Hilbert space.

Our trial wave functions employed in the FT-VMC method flexibly describe the ground states of a wide range of
strongly correlated electron systems from generalized Heisenberg-type quantum spin systems to
multi orbital Hubbard-type Hamiltonians.
The flexibility of the trial wave functions also implies potential applications of the present FT-VMC method to various systems 
at finite temperatures.

Although we have shown that the linear combination of Pfaffian wave functions with many-body correlations
offer a better truncation of the Hilbert space, 
an other efficient way of performing the truncation 
may further improve the truncation.
The tensor networks may be such a candidate,
as examined in a simulation of the transverse Ising model\cite{PhysRevB.86.245101}.
However,  the huge entanglement inherent in the TPQ states at 
high temperatures, which obey the volume law, still hampers the 
direct application of the tensor networks as truncated TPQ states.
An intriguing future issue is
to develop methods of operating the tensor-network projections to the finite-temperature variational 
wave functions treated in this paper  
to overcome this difficulty.

{\bf Acknowledgments}~
The FT-VMC codes in this paper were developed on the basis of the mVMC codes first developed by Daisuke Tahara and Satoshi Morita and extended for complex numbers by Moyuru Kurita. 
We thank Shiro Sakai and Dai Kubota for providing us with QMC data. 
To compute the Pfaffians of skew-symmetric matrices, we employed the PFAPACK~\cite{Wimmer:2012:A9E:2331130.2331138}.
Our benchmark calculations were partly carried out at the Supercomputer Center, Institute for Solid State Physics, University of Tokyo. K. T. and K. I. were supported by Japan Society for the Promotion of Science through the Program for Leading Graduate Schools (MERIT). The authors were financially supported by the MEXT HPCI Strategic Programs for Innovative Research (SPIRE) and the Computational Materials Science Initiative (CMSI). We also thank the support by RIKEN Advanced Institute for Computational Science through the HPCI System Research Project (hp150211).

%
%
\appendix
%
\section{Algorithm and accuracy of FS-TPQ method}
%
In this Appendix, we explain the algorithm of the FS-TPQ method~\cite{PhysRevLett.108.240401}.
We also examine the accuracy of the FS-TPQ method for a small system by
comparing the results of exact diagonalization.

First, we prepare a random state $\ket{\Psi_0}=\sum_{i}c_i\ket{i}$, where $\{\ket{i}\}$ is an orthonormal basis set. Next, we take a constant $l$ that is larger than the maximum
eigenvalue of the Hamiltonian. We need $l$ not only to reach the ground state by the power method
but also to control the speed of convergence. 
We take $l=8.0$ in this calculation.
We start from $\ket{\Psi_0}$ and derive the $k$th state $\ket{\Psi_k}$ by
multiplying repeatedly. In each step, the energy $u_k$ decreases gradually down to the ground-state energy as
\begin{align}
u_k \equiv \frac{\braket{\Psi_k|\hat{h}|\Psi_k}}{\braket{\Psi_k|\Psi_k}},\;\;\;\ket{\Psi_k}\equiv (l-\hat{h})^k\ket{\Psi_0},
\end{align}
where $\hat{h} = \hat{H}/N_s$ is the Hamiltonian per site.
The $k$th state $\ket{\Psi_k}$ is the TPQ state corresponding to a certain inverse temperature $\beta (u_k, N_s)$.
Then,  $\beta (u_k, N_s)$ can be easily estimated by the following relation:
\begin{align}
\beta (u_k, N_s)=\frac{2k}{N_s(l-u_k)}+O(1/N_s).
\end{align}
$\ket{\Psi_k}$ is one of the pure states corresponding to $\beta (u_k, N_s)$; thus, we can obtain the expectation value of the physical value $\hat{A}$ by replacing the ensemble average with $\ket{\Psi_k}$.

\begin{figure}[h!]
\begin{center}
\includegraphics[width=8cm]{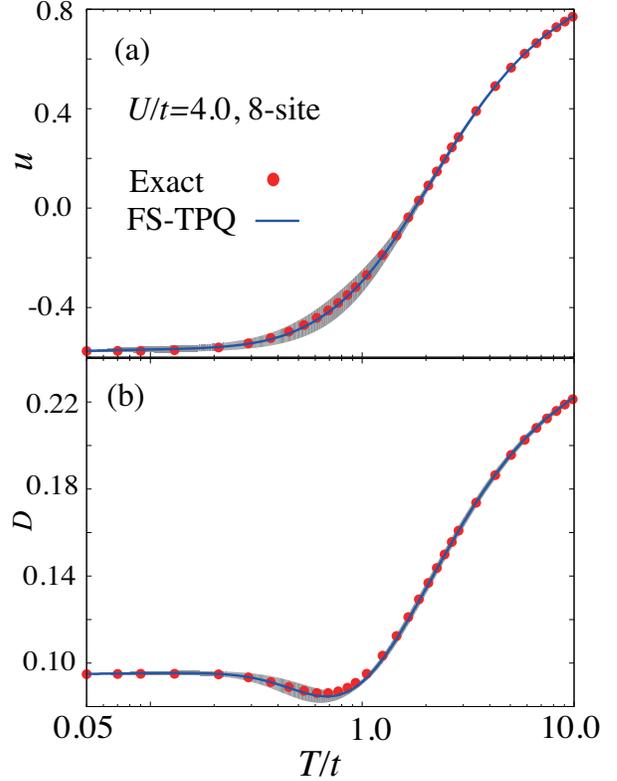}
\end{center}
\caption{\label{fig:8site}(color online)
(a)~Energy per site $u$ calculated from the FT-VMC method
for a 8-site half-filled Hubbard ring with $N_{\uparrow}=N_{\downarrow}=4$ and $U/t=4.0$.
The exact results are shown by the (red) points
in comparison with those of the the FS-TPQ method shown by the  (blue) curve with  (gray) error bars.
(b)~Temperature dependence of double occupancy $D$ per site calculated by FS-TPQ method and the exact diagonalization.
The exact results are shown by the  (red) points
in comparison with those of the FS-TPQ  method shown by the  (blue) curve with  (gray) error bars.
}
\end{figure}

We compare the energy per site $u$ and double occupancy $D$ per site of the FS-TPQ method and those for the exact diagonalization of an 8-site half-filled Hubbard ring with $N_{\uparrow} = N_{\downarrow} = 4$ and $U/t = 4.0$. 
We note that the error bars of the 8-site Hubbard ring are larger than those of the 16-site
Hubbard ring because the error bars are determined
by the dimensions of the Hilbert space~\cite{PhysRevLett.108.240401,PhysRevLett.111.010401}.
As shown in Fig. ~\ref{fig:8site},
we confirm that the FS-TPQ method well reproduces the results of
the exact diagonalization.
 
%
\section{Criterion for accuracy of FT-VMC method}
%
Next, we introduce  the criterion for the accuracy of the FT-VMC method when exact data are not available.
If we consider a trial state $\ket{\bar{\psi}(\alpha(\tau))}$ corresponding to $k_{\rm B}T=1/2\tau$, the {\it exact}
next-step wave function $\ket{{\psi}_{\rm ex}(\tau+\Delta\tau)}$ in the imaginary-time evolution 
is given as
\begin{align}
\ket{{\psi}_{\rm ex}(\tau+\Delta\tau)}=(1-\Delta\tau(\hat{H}-\langle\hat{H}\rangle))\ket{\bar{\psi}(\alpha(\tau))}
\end{align}
if $\Delta\tau$ is small within the Euler method.
On the other hand, we construct the next-step trial wave function using the FT-VMC method as
\begin{align}
&\ket{{\psi}(\alpha(\tau+\Delta\tau))}\nonumber\\
=&\ket{\bar{\psi}(\alpha(\tau))}+\sum_{k}\Delta \alpha_k \ket{ {\partial_{\alpha_k}} \bar{\psi}(\alpha(\tau))},
\end{align}
by employing the TDVP.
Here, we define the overlap $\Delta_{\rm overlap}(\tau)$ between two normalized wave functions as
\begin{align}
\Delta_{\rm overlap}(\tau)\equiv |\braket{{\bar{\psi}}_{\rm ex}{(\tau+\Delta\tau)}|{\bar{\psi}}(\alpha(\tau+\Delta\tau))}|^2,
\end{align}
where $\ket{\overline{\cdots}}$ is defined by Eq.~(5).
$\Delta_{\rm overlap}(\tau)$ ranges within $[0, 1]$ and converges to unity when the imaginary-time evolution is performed exactly. In addition, we define the cumulative value of  $\Delta_{\rm overlap}(\tau)$ as $\Delta_{\rm err}(\tau)$, that is, 
\begin{align}
\Delta_{\rm err}(\tau=l\Delta\tau)\equiv \prod^{l}_{k=0}\Delta_{\rm overlap}(k\Delta\tau).
\end{align}
From the formulation of the TPQ states\cite{PhysRevLett.111.010401}, the partition function $Z(\beta=2\tau, N_s)$ is expressed as
\begin{align}
Z({ \beta=2\tau}, N_s)=L{\braket{\Psi_0|e^{-{2}\tau\hat{H}}|\Psi_0}},
\end{align}
where $\hat{H}$ is the Hamiltonian, $\ket{\Psi_0}$ is a normalized random vector and $L$ is the dimension of the Hilbert space. 
Therefore, $1-\Delta_{\rm err}(\tau)$ expresses the error 
in the FT-VMC result for $Z(\beta=2\tau, N_s)$ estimated from the difference between the two pure states, one  constructed by the FT-VMC method and the other constructed from the TPQ states.
In our calculation, we regard the TPQ state at $T=\infty$ 
as being the same as the present FT-VMC wave function. Since it is not exact, the errors of our FT-VMC method can be larger than $1-\Delta_{\rm err}(\tau)$.
However, the agreement of physical quantities estimated by the present FT-VMC method with the exact values indicates that the additional error is small.

\begin{figure}[h!]
\begin{center}
\includegraphics[width=8cm]{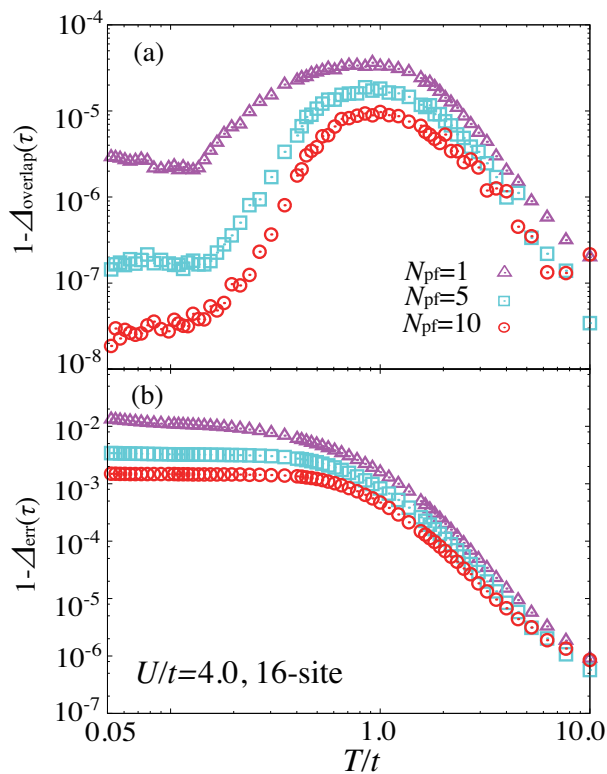}
\end{center}
\caption{\label{fig:ov}(color online)
(a)~ $1-\Delta_{\rm overlap}(\tau)$ calculated from the FT-VMC method 
for a 16-site half-filled Hubbard ring with $N_{\uparrow}=N_{\downarrow}=8$ and $U/t=4.0$.
The results for $N_{\rm pf}=1, 5$, and 10 are shown by (purple) triangles, (light-blue) squares,  and (red) circles, respectively.
(b)~$1-\Delta_{\rm err}(\tau)$ calculated by the FT-VMC method.
The results of $N_{\rm pf}=1, 5$, and 10 are shown by (purple) triangles, (light-blue) squares, and (red) circles, respectively.}
\end{figure}

We calculate and analyze $\Delta_{\rm overlap}(\tau)$ and $\Delta_{\rm err}(\tau)$ for a 16-site Hubbard ring at half-filling ($N_{\uparrow} =N_{\downarrow} =8$ and $U/t=4.0$).
We take the same level of random initial states for $N_{\rm pf}=1, 5,$ and 10 as in Eq.~(27) 
and perform FT-VMC calculations.  
As shown in Fig.~\ref{fig:ov}, $1-\Delta_{\rm overlap}(\tau)$ is similar  to $\Delta u$ in Fig.~\ref{fig:1dU4}. Therefore, we estimate the error of  the FT-VMC method from $\Delta_{\rm overlap}(\tau)$.
In addition,  $1-\Delta_{\rm error}(\tau)$ decreases when we increase the number of Pfaffians (see Fig.~\ref{fig:ov}).
As shown in Fig.~\ref{fig:diff},  the error in $u$ is less than 0.02 and comparable to $\epsilon$ when we take 10 Pfaffians in this system.
The present result suggests that $1-\Delta_{\rm error}(\tau)\approx 10^{-3}$  is required to
achieve this level of accuracy. 

\bibliographystyle{jpsj_mod}
\bibliography{jpsj_final}

\end{document}